# Silicon detectors for the n-TOF neutron beams monitoring

L. Cosentino[1], A. Musumarra[1,2], M. Barbagallo[3], N. Colonna[3], L. Damone[3],
A. Pappalardo[1], M. Piscopo[1], P. Finocchiaro[1]

for the n-TOF collaboration

[1] INFN Laboratori Nazionali del Sud, Catania, Italy
[2] Dipartimento di Fisica, Università di Catania, Italy
[3] INFN Sezione di Bari, Italy

**Abstract**

During 2014 the second experimental area EAR2 was completed at the n-TOF neutron beam facility at CERN. As the neutrons are produced via spallation, by means of a high-intensity 20 GeV pulsed proton beam impinging on a thick target, the resulting neutron beam covers an enormous energy range, from thermal to several GeV. In this paper we describe two beam diagnostic devices, designed and built at INFN-LNS, both exploiting silicon detectors coupled with neutron converter foils containing $^6$Li. The first one is based on four silicon pads and allows to monitor the neutron beam flux as a function of the neutron energy. The second one, based on position sensitive silicon detectors, is intended for the reconstruction of the beam profile, again as a function of the neutron energy. Several electronic setups have been explored in order to overcome the issues related to the gamma flash, namely a huge pulse present at the start of each neutron bunch which may blind the detectors for some time. The two devices have been characterized with radioactive sources at INFN-LNS and then tested at n-TOF. The wide energy and intensity range they proved capable of sustaining, made them quite attractive and suitable to be used in both EAR1 and EAR2 n-TOF experimental areas, where they became immediately operational.

## 1 Introduction

The demand for new and high precision cross section data in reactions induced by neutrons is continuously growing, in order to satisfy the requirements from several fields of science and technology, fundamental and applied research. Several cross section databases exist worldwide, used for a variety of applications, that are in constant need of new and/or better resolution data. This is why several neutron facilities are currently in operation all over the world, with new ones under construction and the European Spallation Source (ESS) on top of all [1].

In the coming years neutron beam intensities never reached so far will allow to open new scientific and technological frontiers. The n-TOF facility, started at CERN in 2001, produces high intensity pulsed neutron beams in a wide energy range (from thermal to few GeV). By means of the time-of-flight technique it provides an energy resolution that also allows to study very narrow resonances. Recently, a second experimental area (EAR2) has been built at n-TOF at a shorter distance from the spallation target, with the aim of making feasible challenging measurements requiring a much higher neutron flux than available in the first experimental area or in most other facilities around the world.

In order to ensure high quality measurements, the neutron beams must be fully characterized as a function of the neutron energy, in particular by measuring the flux and the beam transverse profile. In this paper we describe two suitable beam diagnostic devices we designed and built in order to equip the newly completed second experimental area (EAR2), both based on silicon detectors coupled with converter foils containing $^6$Li [2]. As will be shown in the following, the first device is a set of four silicon pads intended for flux monitoring, whereas the second one is based on a position sensitive silicon detector (either with strips or another technique) intended for x-y-E beam profile reconstruction. The two devices, respectively named SiMon2 and SiMon2D, after being characterized at LNS became operational at n-TOF, showing very nice performances.

We summarize here the basic features of the n-TOF facility, whose full description can be found in refs.[3],[4].

The neutron beams are produced by spallation, by means of a primary proton beam accelerated at 20 GeV by the Proton Syncrotron (PS) hitting a thick lead target with a repetition period around two seconds. The first experimental area (EAR1) is placed at the end of a 180 m long horizontal flight path, whereas EAR2 is located at the end of a 19 m long vertical one. The overall flux in EAR1 is about $10^5$ n/s/cm$^2$ and, due to the ten times shorter flight path, about $10^7$ n/s/cm$^2$ in EAR2. The new experimental area will thus allow one to study processes characterized by very small cross sections, by measuring with a higher S/N ratio at the price of a slightly worse energy resolution with respect to EAR1. Neutron induced reactions of astrophysical interest with highly radioactive targets will also be accessible for measurements.

The devices SiMon2 and SiMon2D were built to measure respectively flux and profile as a function of the neutron kinetic energy. The neutron time of flight as a function of kinetic energy for EAR1 and EAR2 is shown in Figure 1. In the figure one can also see that the time interval needed to collect all of the neutrons produced in each bunch (the slowest being at thermal energy) is respectively about 100 ms in EAR1 and 10 ms in EAR2.

In particular fast neutrons arriving in EAR2 with energy above 1 MeV take less than 2 μs after the gamma flash. Therefore the read-out electronics must be fast enough to get rid of the huge saturated electronic pulse arising in consequence of such a flash, and be ready to quickly restore the signal baseline in order to handle the incoming high energy neutrons.

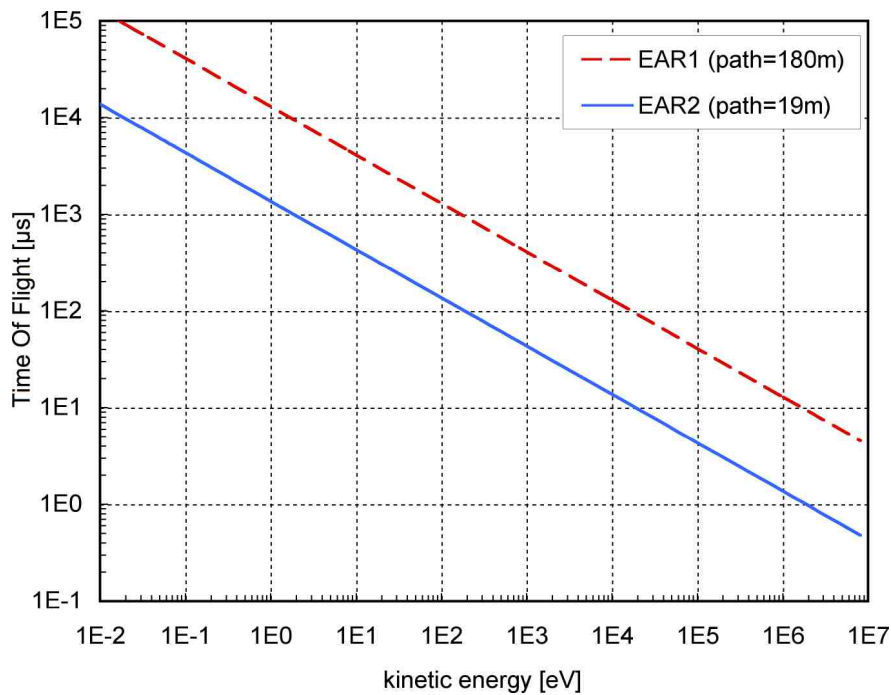

Figure 1. Neutron time of flight as a function of kinetic energy, in EAR1 and in EAR2.

SiMon2 and SiMon2D are housed in an aluminum vacuum chamber, designed at LNS with all the requirements needed to be installed along the vertical beam line in EAR2. This chamber is vacuum tight and was tested down to $10^{-7}$ mbar, even though the vacuum level so far required inside the beam line is only $10^{-2}$ mbar. SiMon2 is permanently installed along the beam line and continuously used during the measurements. SiMon2D can be remotely inserted/removed in the beam path by means of a motor driven actuator, thus intercepting the beam for the time interval required to produce the beam profile. Such an interval can be a few hours or less, depending on the beam intensity and on the device configuration. Both devices were geometrically aligned with the chamber, with an additional fine tuning possible in case of a slight misalignment with respect to the true beam axis.

## 2 The SiMon2 neutron flux monitor

### 2.1 Device description

The device developed for the flux measurement in the new experimental area EAR2 is an upgraded version of SiMon, which has been in operation in EAR1 since more than ten years [5]. SiMon2 has basically kept a similar design, while incorporating a few mechanical and structural improvements making it suitable for being used both in EAR1 and EAR2. It consists of four silicon pad detectors 3cm x 3cm, 300 µm thick (Micron Semiconductors, MSX09-300), arranged around a $^6$LiF neutron converter layer as sketched in Figure 2.

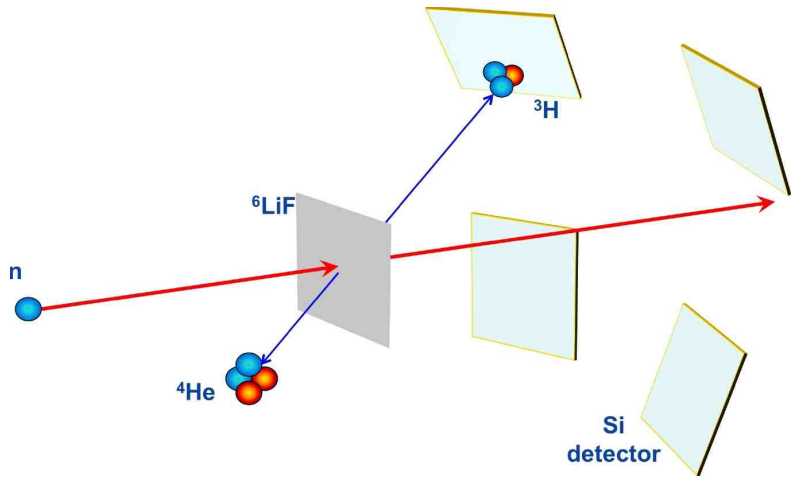

Figure 2. Sketch of the SiMon2 device. Following a neutron capture in $^6$Li, either the triton or the alpha particle from the decay can be detected, representing a signature of the neutron.

The $^6$LiF converter, a rather stable salt, is deposited onto a 1.5 µm thin Mylar foil. The conversion reaction exploited is:

$$^6Li + n \rightarrow ^3H \ (2.73 \ MeV) + \alpha \ (2.05 \ MeV) \tag{1}$$

which is the only possible decay channel following the capture and is free of gamma rays. The spectrum of deposited energy in the silicon detectors has a typical shape, and allows to easily discriminate the capture reaction products from the background gamma rays. In Figure 3 we show a front view of the four silicon detectors of the SiMon2 beam monitor. The $^6$LiF converter foil, to be installed in front of the central square hole, is not present in the picture.

The 300 µm thickness for the silicon detectors was chosen as a trade-off between the gamma sensitivity and the detector capacitance: with this choice the detector capacitance is low enough to guarantee low-noise operations using rather standard preamplifiers even in a noisy experimental environment. At the same time, the maximum energy deposited by gamma rays is below 1 MeV, thus allowing a nice separation from alphas and tritons produced by neutron capture.

The $^6$LiF layer thickness can be chosen depending on the effective beam intensity expected, basically tuning the counting rate to the experimental needs. At LNS several converters of different thickness have been produced, from 400 µg/cm² down to 10 µg/cm², by evaporating $^6$LiF on Mylar foils. The 1.5 µm thickness for the Mylar substrate was chosen in order to ensure the required mechanical stability, while still being thin enough to neglect any appreciable interference in the beam. Moreover, in this range of thicknesses the energy degradation of the produced triton and alpha in the $^6$LiF layer is quite small.

Another possible choice is to deposit the converter on a thicker substrate, and install it with the substrate facing the detectors. In such a way one can stop the alpha particles in the substrate, and rely on the tritons only, thus simplifying the flux reconstruction should the gamma background (or noise) be unexpectedly high.

The main differences with the previous SiMon device are the smaller size of the detectors and the use of $^6$LiF instead of pure $^6$Li as converter. Pure $^6$Li is highly reactive, therefore it had to be enclosed between two thin protective passivation layers typically made from carbon. The production procedure was quite complex, and the final converter very delicate. Conversely, $^6$LiF is quite stable and can be easily deposited in a wide range of thickness onto several different substrates without any protective layer. The thickness of the converter layer can be monitored during the deposition by means of a well known technique based on the frequency variation of a quartz oscillator [6]. The thickness was later checked, along with its uniformity, by means of the energy loss of alpha particles from a source when crossing the converter itself and impinging on a silicon detector placed behind it [2]. The thickness corresponded to the nominal value within 2%, and the non-uniformity of the deposited layer was better than 1%.

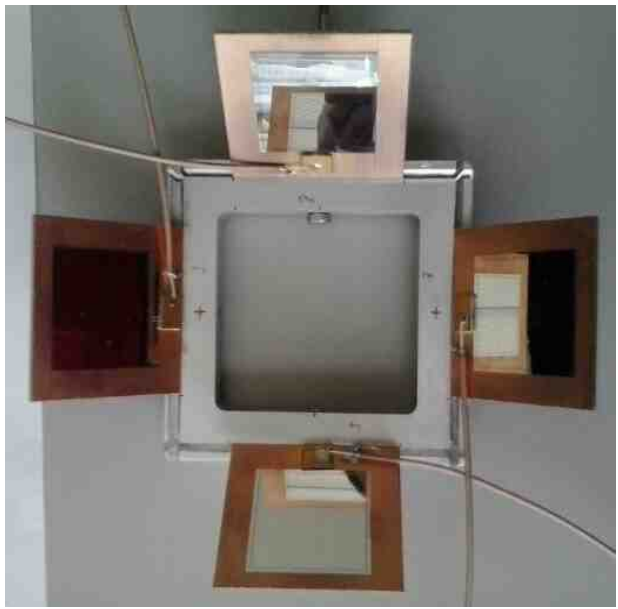

Figure 3. The SiMon2 neutron beam monitor, with its four silicon detectors. The $^6$LiF converter foil (not shown) is to be installed in front of the central square hole at 7 cm distance.

All of the mechanical components, i.e. vacuum chamber, flanges, converter holder, detector holder, are made from aluminum, in order to keep the neutron induced gamma background as low as possible. SiMon2 is assembled on an aluminum flange and then fixed to its vacuum chamber that was pre-aligned to the neutron beam axis. Four vacuum-tight lemo feedthroughs allow to connect the detectors to a charge preamplifier module (produced by NeT-Instruments, 32 channels, 45 mV/MeV sensitivity), operating in air and installed close to the flange. Its outputs are fed into four fast shaping amplifiers featuring an efficient pole-zero restoration (Ortec TFA 474), in order to minimize the duration of the saturation due to the gamma flash. The chosen shaping constants were 100 ns as differentiation time and 200 ns as integration time.

The data from SiMon2 are used for the normalization of data collected with the other detector systems throughout the experiment. Moreover, as the cross section of the reaction (1) is well known (Figure 4), one can also perform an absolute normalization to the number of impinging neutrons as a function of their energy and measure absolute cross sections. In order to do this one has to consider the geometrical acceptance of SiMon2 and the angular distribution of the products of reaction (1) by means of Monte Carlo numerical simulations. At low neutron energy such a distribution is isotropic in the center of mass frame, and in Figure 5 we show how it is transformed into the laboratory frame for three different incoming neutron energies. We can see that even at high energy the kinematic contribution to the observed anisotropy in the lab frame is below 20%. The main contribution to the non-isotropic distribution comes from the dynamics of the reaction itself: the very few existing data, and the evaluation taken from the JEFF-3.1 library, indicate that around the 240 keV resonance the distribution in the C.M. frame is foreseen to become forward-backward peaked, as shown in Figure 6. A forthcoming paper with the detailed description of the flux reconstruction method is in preparation. Just to give a numerical indication, the integrated solid angle seen from the converter foil is about 4.6%. For neutron energy below 1 keV, hence isotropic distribution, and due to the back-to-back emission of the two particles, the SiMon2 geometrical acceptance of a neutron signature becomes about 9.2%.

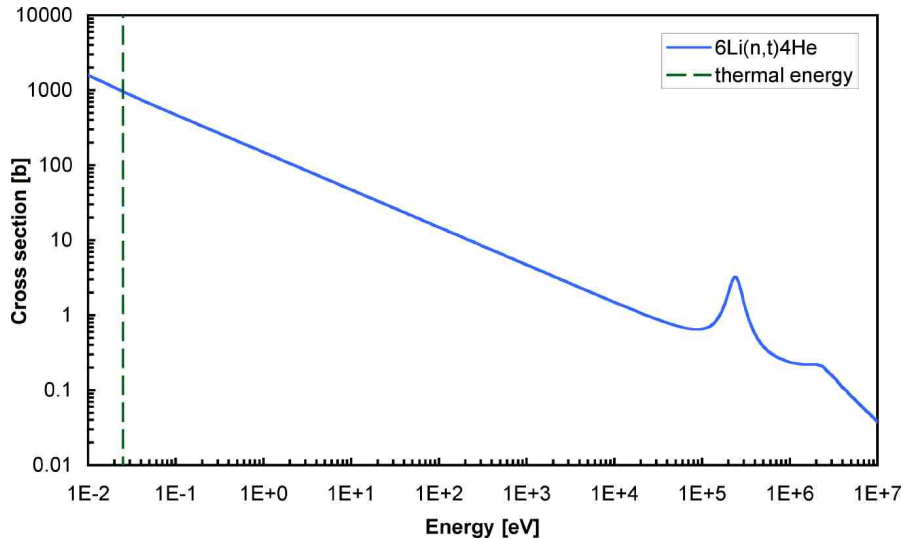

Figure 4. Cross section of the reaction $^6$Li(n,t)$^4$He.

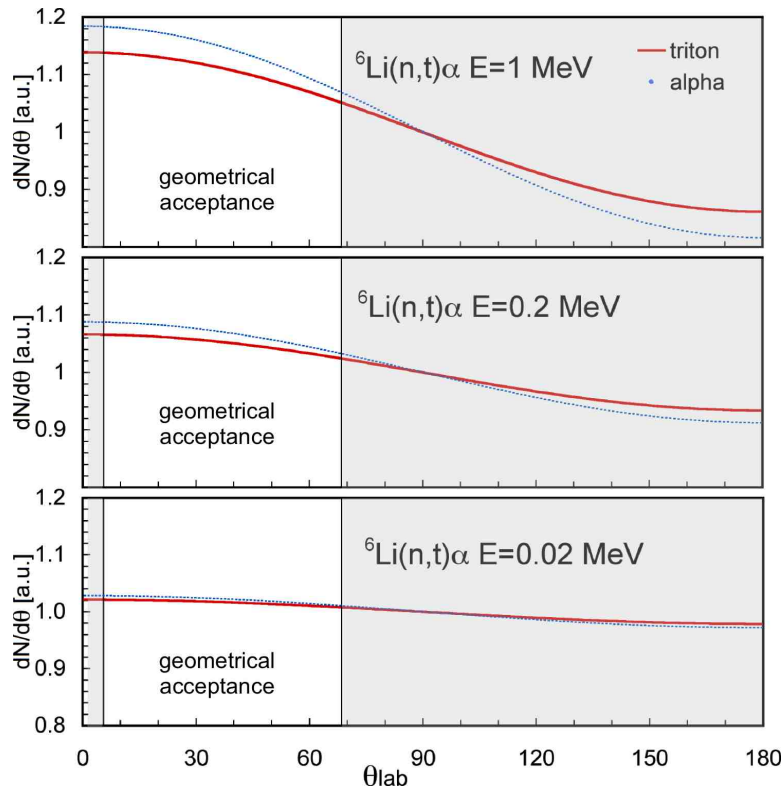

Figure 5. The C.M. isotropic distribution of tritons and alphas as transformed into the laboratory frame, for three different incoming neutron energies. The rough geometrical acceptance of SiMon2 is highlighted.

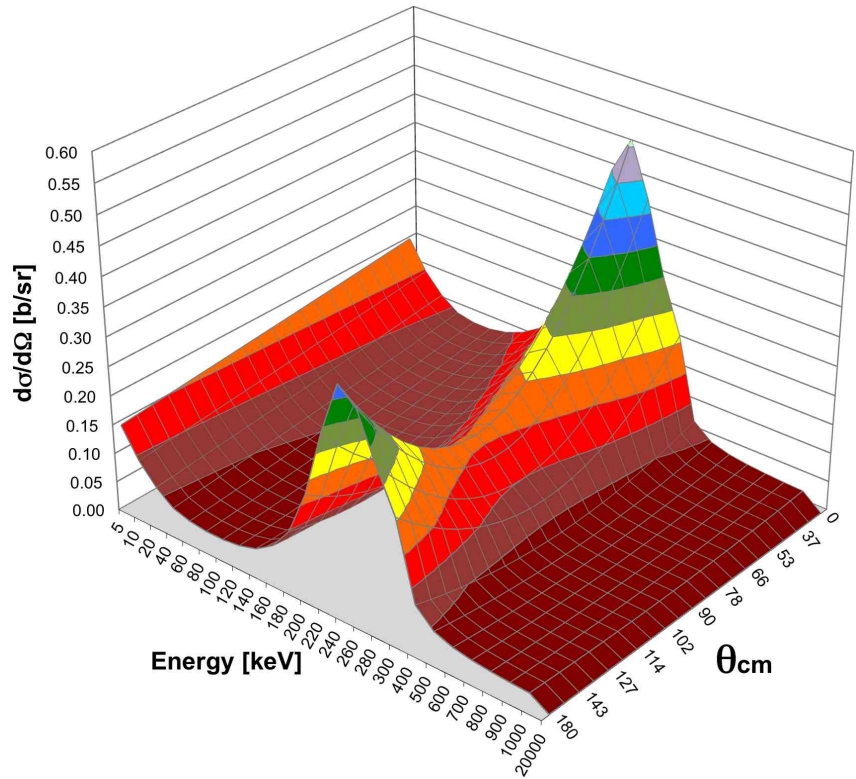

Figure 6. The expected angular distribution of tritons and alphas from reaction (1) in the C.M. frame for several incoming neutron energies. Around the 240 keV resonance the distribution is foreseen to become forward-backward peaked.

*2.2 Bench test*

Before the installation in beam of the SiMon2 flux monitor, several tests were performed with its converter/detectors. Each of the four detectors was first calibrated in vacuum with a Pu-Am-Cm source, which emits alpha particles with respective average energies of 5.147, 5.477 and 5.793 MeV. The measured energy resolution ($\approx$ 30 keV FWHM) was well below 1%, as testified by the evidence in the spectrum of the secondary satellite peaks due to alpha decays from channels with lower branching ratios, whose energy differences from the main peaks range between 12 and 40 keV (Figure 7).

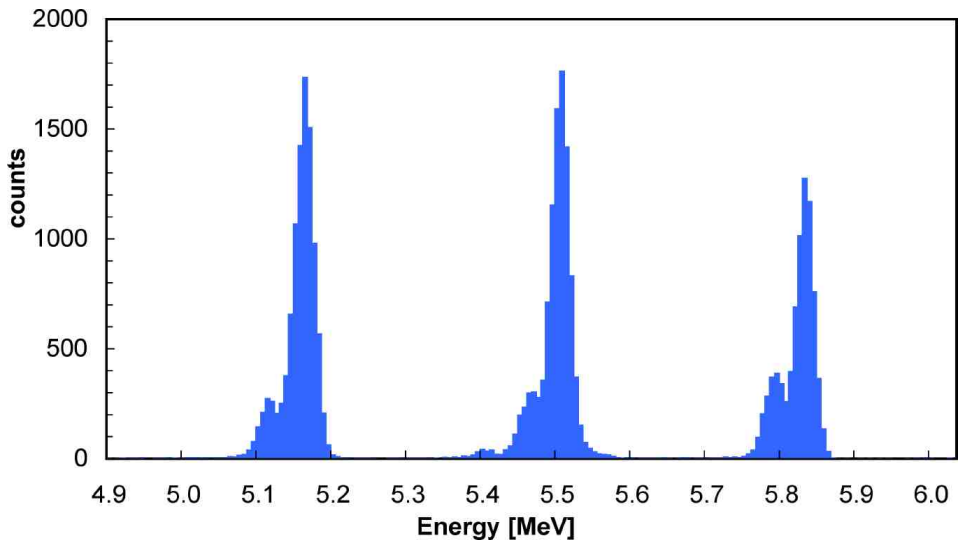

Figure 7. Energy calibration spectrum of one of the SiMon2 silicon detectors. The employed alpha source was a mix of Pu-Am-Cm. The width of the peaks ($\approx$ 30 keV FWHM) and the evidence of the satellite peaks (at slightly lower energy levels) testify a resolution well below 1%.

The following test was performed by placing a neutron converter on top of each detector, and then exposing them to an AmBe neutron source (≈ 1.6E6 n/s), suitably moderated with $CH_2$ slabs, and to a $^{60}$Co gamma source. This test was done in air due to logistic reasons, as we had no vacuum chamber small enough to be placed near the neutron source inside the moderator. The resulting spectra, proving the detector sensitivity to neutrons and the gamma/neutron discrimination capability, are shown in Figure 8.

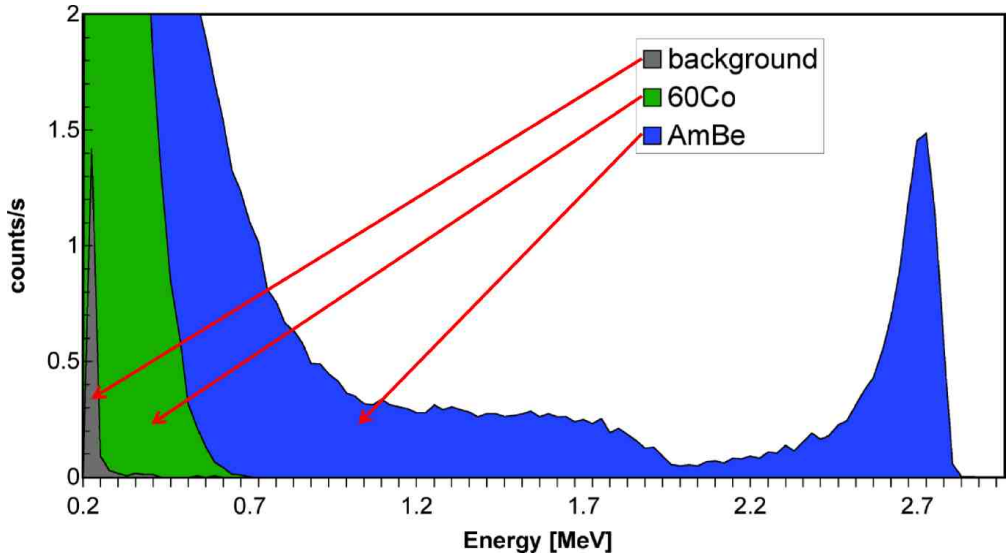

Figure 8. Energy spectrum of one of the SiMon2 silicon detectors when exposed respectively to: (i) environment background; (ii) a $^{60}$Co gamma source; (iii) the AmBe neutron source.

## 2.3 In-beam tests

The thickness of the $^6$LiF converter layer to be used in SiMon2 should provide a reasonable counting rate in the silicon detectors and, at the same time, a negligible probability of double hit. Indeed, a significant amount of double hits could seriously affect the reliability of the flux measurement. The optimal thickness, considering the high flux at the second experimental area of n-TOF, was found to be around 100 μg/cm$^2$. With this thickness the pile-up probability is small, while a reasonable statistics on the neutron flux can still be obtained in runs of a few hours. The bias voltage on the four detectors was 20 V, sufficient to operate in full depletion regime, and the corresponding reverse current was about 12 nA.

To avoid problems related to the large prompt signal (the so-called gamma flash), a fast electronics listed in section 2.1 minimizes the saturation of the prompt signal, thus allowing flux measurements up to 1 MeV in EAR2 and to 10 MeV in EAR1. We are currently working to a further improvement of the electronics, in order to shift these energy limits upwards.

Once the electronics was tuned, the amplitude spectra of the silicon detectors showed a very clear identification of the reaction products and a clean separation from the electronic noise. A sample spectrum for one of the detectors, collected in a six minute run at EAR2, is shown in Figure 9. An impressive separation is visible between tritons, alphas and background gamma rays, at variance with Figure 8 where the separation is worse due to: (i) operation in air, that produces an energy spread towards lower energies; (ii) a huge background, produced by neutrons and gammas from the AmBe source, directly hitting the detector. The energy spread of the alpha and triton peaks observed in Figure 9 is only due to the thickness of converter layer crossed by these particles following their creation by the neutron interaction with a $^6$Li nucleus. As the alpha particle is heavier and has a lower kinetic energy than the triton, its spread is larger.

In Figure 10 we show a sample scatter plot of deposited energy versus neutron energy. Up to a few keV neutron energy each vertical slice of the plot is a replica of Figure 9 with different statistics.

Around 240 keV the enhancement due to the $^6$Li cross section resonance (see Figure 4) is clearly visible. At the same time the overall shape bends upwards, as the kinematical boost due to the impinging neutron energy, and the consequent motion of the center of mass, increases the kinetic energy of the detected particles in the lab frame. By selecting the region with alphas and tritons only, the number of counts in each vertical slice corresponds to the number of detected neutrons with kinetic energy in a given interval.

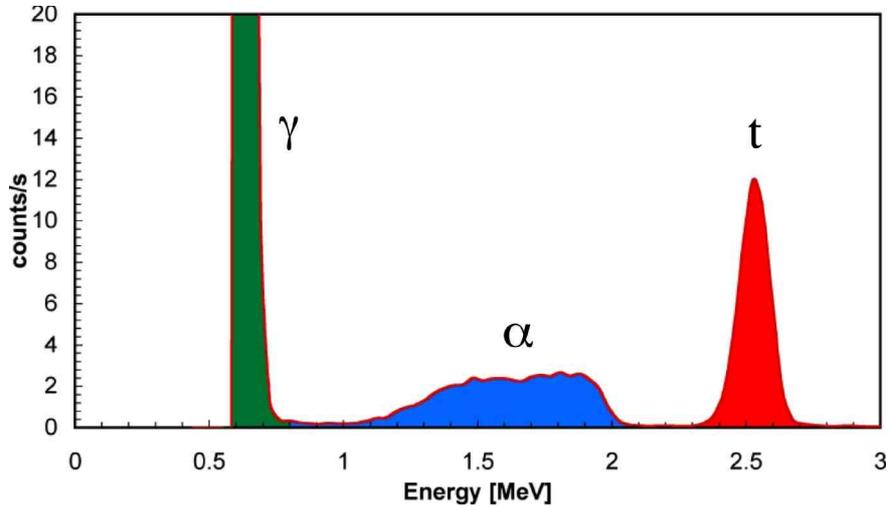

Figure 9. A sample spectrum for one of the SiMon2 detectors, collected in a six minute run at EAR2. A clear separation is visible between tritons, alphas and background gamma rays.

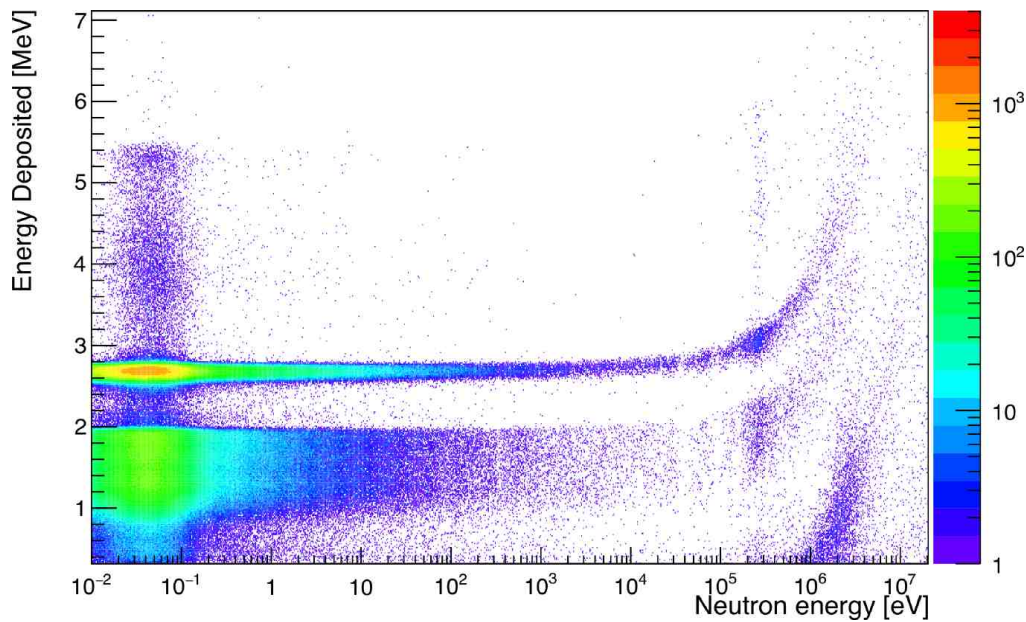

Figure 10. A sample scatter plot of energy deposited in a silicon detector of SiMon2 versus neutron energy. The upwards bend at high neutron energy is due to the kinematical boost. See text for details.

As for the radiation damage, we monitored the reverse current during many days of continuous operation and verified that it kept its 12nA value quite stably. This indicates that SiMon2 is not affected by significant radiation damage and can be installed and used permanently in the high flux environment of EAR2.

## 3  The neutron beam profile monitor: SiMon2D

### 3.1  Device description

Before starting an experiment, and from time to time during it, one needs to check the beam alignment and its spatial profile in the transverse plane. Moreover, as the neutron beam is not monoenergetic, one would like to check these observables as a function of the neutron energy. This is the reason why a beam profile monitor is needed.

Such a device should be able to produce the required information in a short time as compared to the duration of the experiment: as a consequence it is interceptive (i.e. it perturbs the beam) and has to be inserted when needed and then removed. Being interceptive also implies that radiation damage can be a relevant issue to be taken seriously into account.

The beam profile monitor SiMon2D is based on a 5cm x 5cm position sensitive silicon detector coupled to a $^6$LiF converter layer. When in operation the detector faces the beam axis perpendicularly, in order to intercept the full neutron beam. The spatial distribution of triton and alpha particles, produced by the interaction of the neutrons with the converter and hitting the silicon detector, is strongly correlated to the distribution of neutrons in the plane perpendicular to the beam direction. Indeed, being the converter closely coupled with the detector, the proximity focusing introduces a spatial spread of the same order of magnitude of the converter-detector distance (2.5 mm). The current version of SiMon2D makes use of a 16 x + 16 y strip double sided silicon detector, each strip 3 mm wide, 300 μm thick, therefore the space resolution is dominated by the position resolution of the detector (Figure 11). A 400 μg/cm$^2$ $^6$LiF converter layer, deposited onto a 1 mm thick carbon fibre plate, is assembled on top of the detector also providing mechanical protection to the detector. The detector+converter sandwich is fixed to a rod that can be moved on/off beam by means of a motorized actuator operating in vacuum (Figure 12). Two suitable vacuum-tight feedthrough connectors allow to transport the 32 detector signals out of the beam pipe, which are then connected to 32 preamplifier (two Mesytech MPR-16LOG modules) and 32 spectroscopy amplifier (two CAEN N568B module) channels.

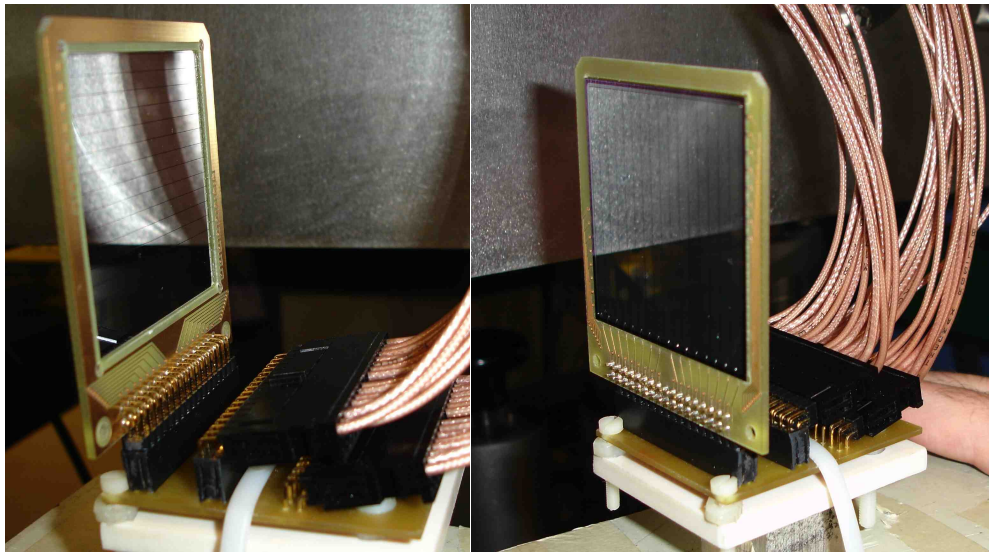

Figure 11. The 16 x + 16 y strip double sided silicon detector employed for the SiMon2D beam profile monitor.

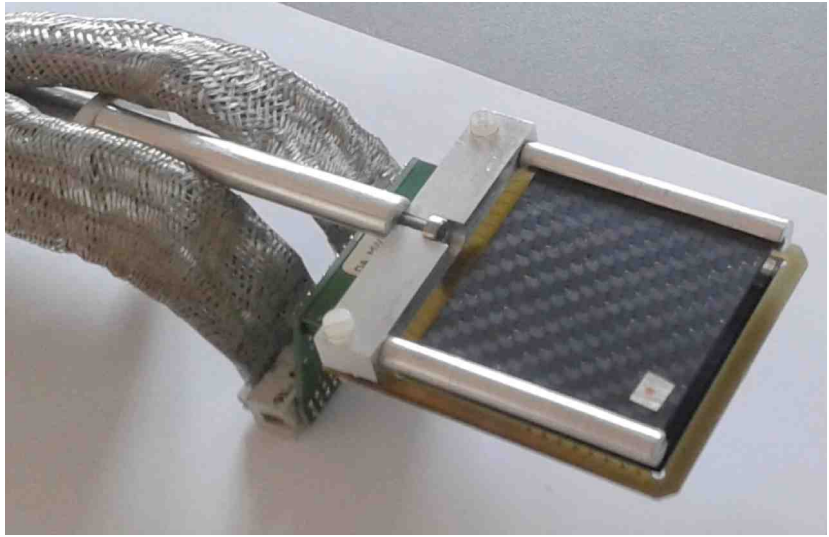

Figure 12. The SiMon2D beam profile monitor assembled on its actuator. The $^6$LiF converter, facing the silicon strip detector, is deposited on a carbon fibre plate.

## 3.2 Bench tests

The first test we performed was aimed at characterizing the silicon strip detector, therefore we installed the bare detector in a vacuum chamber and placed a collimated Pu-Am-Cm alpha source for three minutes in front of each strip. In Figure 13 we show a sample plot of the energy spectra for 16 strips with the three expected peaks. Afterwards we had to verify that the detector could effectively reconstruct profiles. To this purpose we placed the same source a few centimeters in front of it and required the coincidence between one strip on the front side and one in the back, along with a suitable threshold in amplitude, and built the x-y distribution shown in Figure 14. This is exactly what we expected, and therefore we proceeded with a more stringent test involving the detection of neutrons.

We assembled the neutron converter described in section 3.1 on top of the detector but, instead of following the setup shown in Figure 12, we placed the converter as sketched in Figure 15 left. Afterwards we exposed this configuration of SiMon2D to the moderated AmBe neutron source and, after applying the same front/back coincidence algorithm, we obtained the x-y plot shown in Figure 15 right.

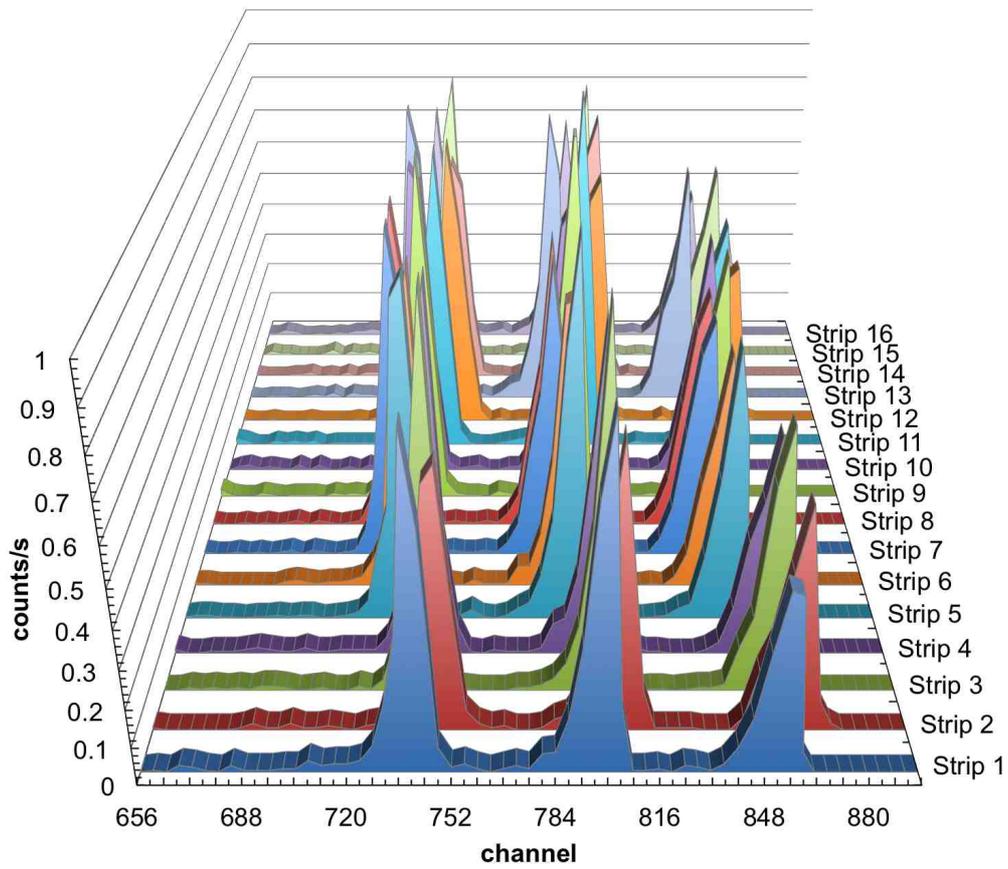

Figure 13. Sample plot of the energy response of 16 strips of SiMon2D to a Pu-Am-Cm alpha source. The three-peak structure is evident in all the strips as expected.

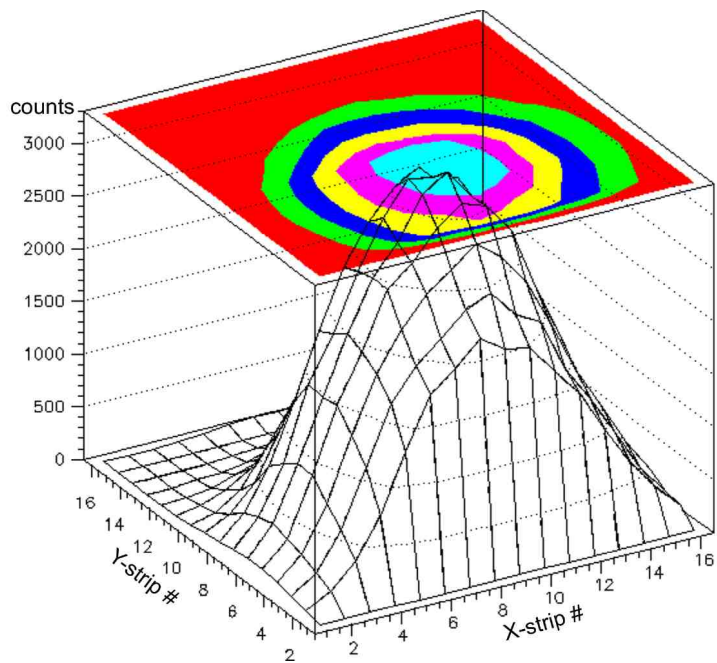

Figure 14. Image of an alpha source placed in vacuum in front of the x-y double-sided silicon strip detector of SiMon2D.

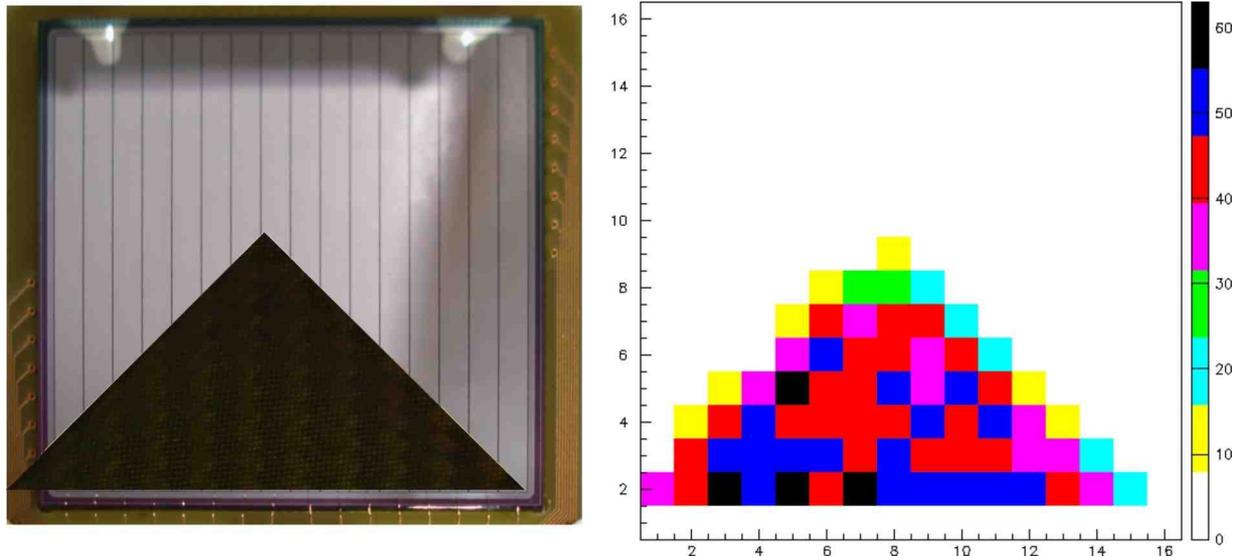

Figure 15. Bench test of SiMon2D with a moderated AmBe neutron source. Left: sketch of a test configuration with only a triangular converter installed on top of the silicon strip detector. Right: the resulting x-y plot.

## 3.3 In-beam tests

During the commissioning of EAR2 SiMon2D was tested as well, in order to evaluate its response in a real neutron beam environment. The device was installed along the vertical beam line of EAR2, housed in the same aluminum vacuum chamber of SiMon2 where the two detectors can coexist, installed in two different branches (Figure 16). Unfortunately at the time of the test the chamber could not be placed under vacuum, and therefore the 2.5 mm distance between the converter and the detector introduced an additional energy degradation for the alphas. On top of this, being the detector in the beam, the low energy background was surely larger than in the case of SiMon2. All this implied that we did not expect to efficiently separate alphas from the background, therefore we were prepared to use only tritons as neutron detection signature.

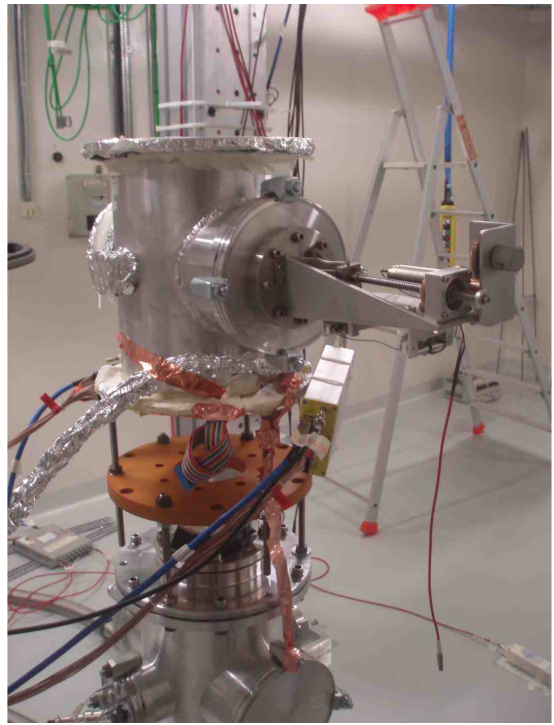

Figure 16. SiMon2D inside its chamber on top of the 19 m vertical beam line during EAR2 installation. The motorized actuator and the moving rod are visible on the right.

The response of all the strips was roughly equalized by adjusting the gain on the two 16-channel spectroscopy amplifiers. A typical energy spectrum for the 16 X-strips is shown in Figure 17, with a well defined triton peak and the alpha peak not separated from the background, as expected.

By requiring the coincidence between one strip on the front side and one in the back, along with a threshold in amplitude selecting only the triton peak, we can build the x-y neutron distribution (i.e. the beam profile). An example of such a profile is shown in Figure 18, inclusive of the whole range of neutron energies. By selecting intervals of time of flight several profiles can be built for different neutron energy windows, but this is the subject of a forthcoming paper currently in preparation.

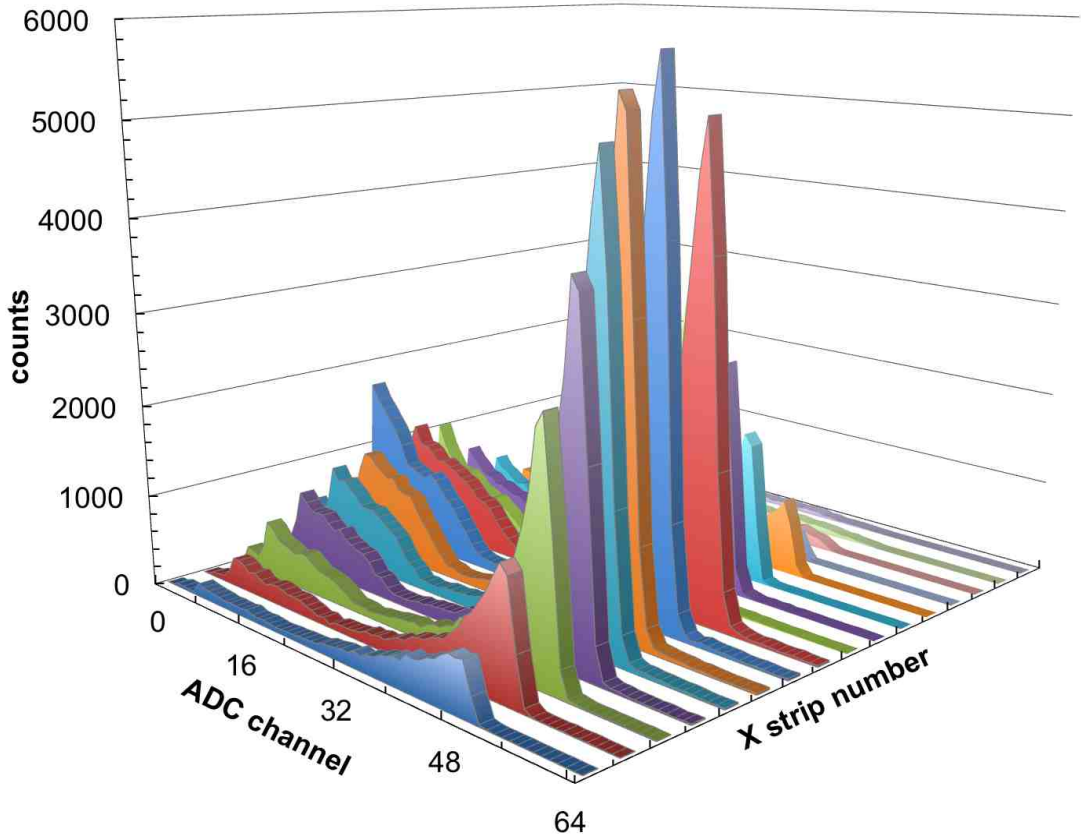

Figure 17. Global energy spectrum for 16 strips of SiMon2D in the neutron beam at EAR2. The triton peak is well defined in all the strips as expected.

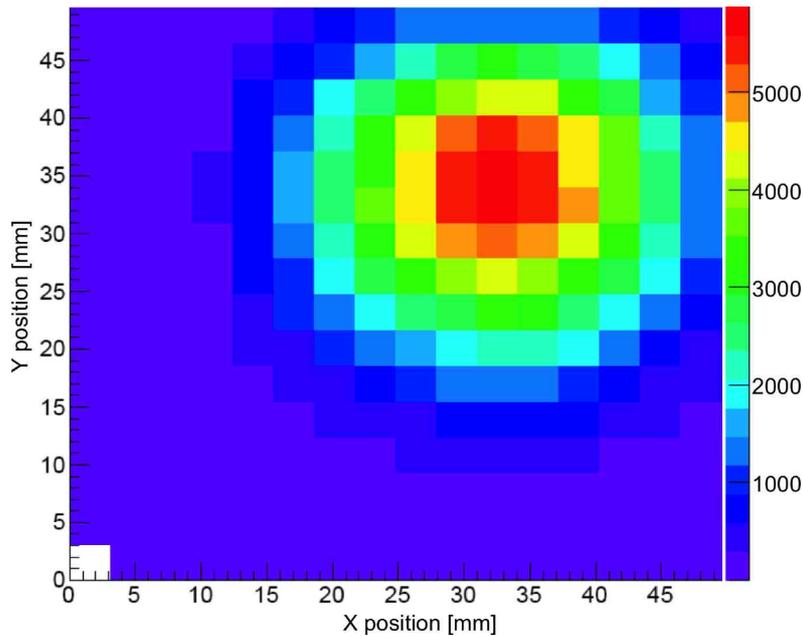

Figure 18. Example of measured neutron beam profile inclusive of all the neutron energies. By selecting intervals of time of flight, several profiles can be built for different neutron energy windows.

The SiMon2D radiation hardness was checked by monitoring the growth rate of the reverse current, that is correlated to the damage in the silicon crystal lattice. By setting a bias of 20 V the initial reverse current was 120 nA with an increase rate of about 10 nA/h when the detector is exposed to the full intensity neutron beam. Even though this is a relevant value, this rate does not represent a serious limitation for the detector performance, as it is capable of operating reasonably up to a reverse current of several microamps. Indeed this corresponds to a countinuous irradiation of few hundred hours, and as the time needed to acquire a profile is typically about one hour, the same detector can be used in SiMon2D to acquire hundreds of profiles without the need of replacing it (i.e. several years).

*3.4 Forthcoming developments*

A drawback of the strip detectors (like for other segmented detectors) is represented by the number of readout channels needed for the data acquisition of all the strips. An alternative solution to reduce the number of channels could be pursued by adopting a resistive partition connecting all the strips on each side of the detector. This reduces the number of channels from 32 down to 4, as the total charge collected event by event in a fired strip is shared by two resistive branches. The ratio between the charge at the two ends (i.e. between the two pulse amplitudes) depends on the strip hit by the particle, and can basically assume only 16 values each one identifying one strip. With this technique the horizontal and the vertical impact position of each detected particle can be identified. A preliminary test of this configuration showed that the solution works, provided that the overall counting rate is low enough to prevent double hits (two or more strips fired simultaneously).

Another alternative solution we are pursuing for a next version of SiMon2D is based on a Position Sensitive Silicon Detector (PSSD): it is a large detector with a special thin resistive layer deposited onto its junction side. The position of each hitting particle is calculated by measuring the fraction of charge collected on the detector corners, while the deposited energy information is obtained from the back side electrode. We are currently testing a sample of a large PSSD (63mm x 63 mm MSPSD-TL-63, produced by Micron Semiconductors) that we plan to install soon.

Even in this case the advantage in the reduction from 32 to 5 channels has the drawback of limiting the maximum allowed counting rate. By suitably reducing the neutron converter layer thickness one can lower the counting rate, even though the exposure time to obtain a reasonable profile increases correspondingly along with the radiation damage. However, the price of such a detector is much lower than that of a strip detector, therefore it could represent a tradeoff between complexity, duration and cost.

## 4  Conclusion

Two detectors for neutron beam monitoring, namely SiMon2 and SiMon2D, were successfully built and tested at n-TOF neutron beam facility during the commissioning of the newly built second experimental area EAR2. Both devices, based on $^6$LiF as neutron converter, have shown such nice performance that they were immediately proposed also for the first experimental area (EAR1).

SiMon2 has shown very promising capabilities both for the beam stability monitoring and also for the differential flux reconstruction as a function of the neutron energy, in light of its clean neutron identification and of the time of flight measurement.

SiMon2D has proved that a reconstruction of the neutron beam profile can be done, and that separate profiles can be built as a function of the neutron energy window, again due to the time of flight measurement capability along with the clean neutron identification.

New developments are currently under way, as the technique of neutron detection by means of $^6$LiF converter foils coupled to solid state detectors looks very promising, in light of the ease of production and of the outstanding results so far obtained.

## 5  Acknowledgments

We are grateful to our friend and colleague Marco Ripani, responsible of the INFN-Energy project, for the constant support to our developments in neutron detection techniques. We are also strongly indebted with Carmelo Marchetta and Eugenio Costa, whom we firmly rely on for the production of the $^6$LiF converters.